\documentclass[pra,twocolumn,showpacs]{revtex4}
\usepackage{graphics}
\newcommand{\tr}{\mbox{Tr}}
\newcommand{\ket}[1]{| #1 \rangle}
\newcommand{\bra}[1]{\langle #1 |}
\newcommand{\inpr}[2]{\langle #1 | #2 \rangle}
\begin{document}

\title{Inequalities That Test Locality in Quantum Mechanics}
\date{\today}
\author{Dennis Dieks}\email{dieks@phys.uu.nl}
\affiliation{Institute for the History and Foundations of Science\\
Utrecht University, P.O.Box 80.000 \\ 3508 TA Utrecht, The
Netherlands}

\begin{abstract}
Quantum theory violates Bell's inequality, but not to the maximum
extent that is logically possible. We derive inequalities
(generalizations of Cirel'son's inequality) that quantify the
upper bound of the violation, both for the standard formalism and
the formalism of generalized observables (POVMs). These
inequalities are quantum analogues of Bell inequalities, and they
can be used to test the quantum version of locality. We discuss
the nature of this kind of locality. We also go into the relation
of our results to an argument by Popescu and Rohrlich (Found.\
Phys.\ {\bf24}, 379 (1994)) that there is no general connection
between the existence of Cirel'son's bound and locality.
\end{abstract}

\maketitle
\section{Introduction}\label{introduction}

The violation of Bell's inequality by the predictions of quantum
theory (both in its non-relativistic and relativistic versions)
shows that quantum theory is non-local in the sense that its
results cannot be reproduced by a hidden-variables theory in which
measurement results depend only on the local settings of the
measuring devices and on the properties of the objects being
measured (a local hidden-variables theory). However, the maximum
violation of Bell's inequality allowed by quantum theory is less
than the maximum violation that is logically possible: quantum
theory obeys Cirel'son's inequality. One might surmise that this
is due to the fact that quantum theory does not abandon locality
completely: after all, in situations of the
Einstein-Podolsky-Rosen (EPR) type the measurements performed on
one wing do not influence expectation values on the other wing
(the no-signaling theorem; in the relativistic context this is the
feature of relativistic causality). Perhaps compliance with a
no-signaling demand restricts the extent to which Bell's
inequality can be violated, and perhaps inequalities like
Cirel'son's can be regarded as a touchstone of this kind of
locality (in the same way as Bell's inequality is a touchstone for
locality in the classical sense).

In this paper we show that this hypothesis is right: the fact that
quantum theory does not violate Bell's inequality to the maximum
logically possible extent is due to features of locality that are
built into the theory. We derive a set of inequalities, and a
strongest inequality representing this whole set, that can be
regarded as quantum versions of Bell's inequality. To make the
analogy with Bell inequalities clear we will analyze how locality
is implemented in quantum theory, and in what sense the quantum
theoretical inequalities we derive are based on locality
assumptions. We will discuss how this relates to a result of
Popescu and Rohrlich \cite{popescu and rohrlich} that at first
sight seems to show that the existence of Cirel'son's bound is
unconnected with locality issues.

\section{Cirel'son's inequality}\label{cirelson}

Consider a probability space in which there are four stochastic
functions, $A, a, B, b$, each of which can take the values $+1$ or
$-1$. The quantity $AB + Ab + aB - ab = A(B + b) + a(B-b)$ can
only be $+2$ or $-2$, from which it follows that the absolute
value of its expectation value is smaller than $2$:
\begin{equation}
|\langle AB + Ab + aB - ab\rangle| \leq \; \langle |AB + Ab + aB -
ab|\rangle =2 \label{bell}.
\end{equation}
This is the form of Bell's inequality that we will consider. The
inequality is respected by physical quantities in classical
theories, as long as these quantities can be represented by
(stochastic) functions on one state space, with a joint
probability distribution---which is ordinarily the case. We will
discuss the connection with locality in sect.\ \ref{locality}.

In quantum mechanics physical magnitudes are not represented by
stochastic functions on a phase space, but by Hermitian operators
on a Hilbert space. Let us now use $A$, $a$, $B$, $b$ to denote
such operators that have eigenvalues $+1$ and $-1$, and let us
consider a combination of them that is analogous to the
combination of quantities in Bell's inequality: $AB + Ab + aB -
ab$, where we assume that the operators occurring in a product
commute. As was first shown by Cirel'son \cite{cirelson}, the
modulus of the quantum mechanical expectation value of this
expression is bounded by $2\sqrt{2}$: $|\langle AB + Ab + aB -
ab\rangle| \leq 2\sqrt{2}$---the upper bound can be attained, as
shown by the example of the singlet state. So Bell's inequality
can be violated by quantum theory; but the quantum expectation
value stays well below the logically possible upper bound of the
expression $|\langle AB + Ab + aB - ab\rangle|$, namely $4$.

Cirel'son's inequality can be proved elegantly by observing
\cite{landau} that if $A^{2}=a^{2}=B^{2}=b^{2}=\openone$ and
$[A,B]=[A,b]=[a,B]=[a,b]=0$, then \[ C^{2}\equiv(AB + Ab + aB -
ab)^{2}= 4\openone - [A,a][B,b].\] It follows from this that
\[\langle C\rangle^{2} \leq \langle C^{2}\rangle \leq ||C||^{2}
\leq 4 + 4||A|| \, ||a|| \, ||B|| \, ||b||= 8, \] or \[|\langle
C\rangle|\leq 2\sqrt{2} \label{land}.\]

An alternative simple proof, which is analogous to the above proof
of Bell's inequality (\ref{bell}) and similar to proofs of other
inequalities that we will give in sections \ref{povms} and
\ref{stronger}, goes as follows.

For a normed state vector $\ket{\psi}$, put $A\ket{\psi}\equiv
\ket{A}$, $B\ket{\psi}\equiv \ket{B}$, $a\ket{\psi}\equiv \ket{a}$
and $b\ket{\psi}\equiv \ket{b}$. Each of these four vectors has a
norm that is $\leq 1$. We now have
\begin{eqnarray}
|\langle C\rangle|& = & |\bra{\psi} C \ket{\psi}| \nonumber \\ & =
& |\inpr{A}{B + b} + \inpr{a}{B - b}| \nonumber \\ &\leq& ||\,
\ket{B} + \ket{b}\, || + ||\, \ket{B} - \ket{b}\, || \nonumber
\\
& \leq & \sqrt{2(1+ \mbox{\bf Re}\inpr{B}{b})} + \sqrt{2(1-
\mbox{\bf Re}\inpr{B}{b})} \nonumber \\ &\leq& 2\sqrt{2}
\label{cirel}.
\end{eqnarray}

The difference between this derivation and the derivation of
Bell's inequality is that for \emph{numbers} $B$ and $b$ with norm
$\leq 1$ we have $|B+b| + |B-b| \leq 2$, whereas for
\emph{vectors} with norm $\leq 1$ we find $||\ket{B}+\ket{b}|| +
||\ket{B}-\ket{b}|| \leq 2\sqrt{2}$. In the latter case the
maximum is attained when $\ket{B}$ and $\ket{b}$ are
perpendicular.

In derivation (\ref{land}) the essential premise is that the
operators ${A,a}$ commute with ${B,b}$. At first sight, derivation
(\ref{cirel}) does not make use of this premise. This impression
is deceptive, however. The operator products occurring in $C$ are
hermitian operators (and therefore representations of physical
quantities) if and only if the operators that are multiplied
commute, and this leads to exactly the same commutativity
requirement as in (\ref{land}). One physical consequence of this
commutativity requirement is that the operators $A$ and $a$ are
jointly measurable with the operators $B$ and $b$. Moreover, it
follows from the commutativity that it does not make any
difference for the expectation values of the operators $B$ or $b$
whether they are measured together with $A$ or $a$ (the
no-signaling theorem). Within the framework of the orthodox
measurement formalism co-measurability and causal independence (in
the sense of no signaling) therefore go together: they both hold
if and only if the commutativity requirement is satisfied. In this
case Cirel'son's inequality also holds.

\section{Generalized measurements}\label{povms}

Above we followed the orthodox point of view about the
mathematical representation of physical quantities in quantum
theory, namely that physical quantities are represented by
hermitian operators. Within this framework joint measurability is
equivalent to commutativity (which in turn leads to the
no-signaling theorem in the context of the EPR experiment). But
there is a more general treatment of measurements in quantum
theory, first developed by Ludwig \cite{ludwig} and Davies
\cite{davies}, in which physical quantities correspond not to
single operators but to collections of positive operators $M_{i}$
on the Hilbert space, such that
\[ M_{i} \geq 0, \;\; \sum_{i}M_{i}= \openone.\]
If the possible outcomes of a measurement of the considered
quantity are ${m_{i}}$, the probabilities of obtaining these
values in a state $\rho$ of the system are given by $\tr \rho
M_{i}$. The mapping $m_{i} \rightarrow M_{i}$ is a positive
operator valued mapping (POVM), representative of the associated
physical quantity $\cal M$.

Two physical quantities $\cal A$ and $\cal B$, represented by sets
of positive operators $\{A_{i}\}$ and $\{B_{j}\}$, respectively,
are jointly measurable if there is a third quantity $\cal O$,
represented by $\{O_{k}\}$, such that
\begin{equation}
A_{i}= \sum_{k\in K_{i}}O_{k}, \;\;\; B_{j}= \sum_{k\in
K^{\prime}_{j}}O_{k},\label{joint}
\end{equation} where $\{K_{i}\}$ and $\{K^{\prime}_{j}\}$ are two
partitions of the index set through which $k$ runs.

If there is an $\cal O$ satisfying Eq.(\ref{joint}) we can measure
it, and infer information about the outcomes and their
probabilities of both $\cal A$ and $\cal B$ by grouping together
the results according to the two partitions. An important feature
of this formalism is that commutativity of the two generalized
observables $\cal A$ and $\cal B$ (in the sense that
$A_{i}B_{j}=B_{j}A_{i}$ for all $i,j$) is a sufficient but not a
necessary condition for their joint measurability. \emph{If} $\cal
A$ and $\cal B$ commute, the products $A_{i}B_{j}$ are positive
operators characterizing the joint measurement of $\cal A$ and
$\cal B$. But in general a joint measurement need not correspond
to product operators (see for a critical analysis of the
significance of these results \cite{uffink0}).

So in the EPR situation we could imagine a joint measurement of
two non-commuting generalized observables $\cal A$ and $\cal B$,
each pertaining to a different wing of the experiment. In this
case it would no longer be true that the mere requirement of
compatibility leads to causal independence (no-signaling), the
product form of the joint measurement operators, and the validity
of Cirel'son's inequality.

However, Busch and Singh \cite{busch} have shown for the EPR
situation, treated by means of the POVMs formalism, that if the
possible values and their probabilities of the quantity measured
at one wing are required to be independent of which quantity is
measured at the other side, the operators representing the
generalized observables at one wing must commute with those at the
other. It follows that in this case the operators corresponding to
the joint measurement take on the product form again. So within
the generalized measurements framework commutativity and product
form are consequences of locality, in the sense of the
impossibility of signaling.

If the measurements in the EPR experiment are represented by
generalized observables, and if locality in the sense of
impossibility of signaling is assumed, Cirel'son's inequality can
again be derived. To see this, consider one pair of the four pairs
of observables, $\cal A$ and $\cal B$, say. Because of the
no-signaling requirement, the corresponding positive operators
$A_{i}$ and $B_{j}$ commute, and the joint measurement of $\cal A$
and $\cal B$ can be represented by four positive operators
$A_{i}B_{j}$, with $i,j = \pm 1$. The expectation value of the
outcomes of this joint measurement, in the pure state
$\ket{\psi}$, becomes:
\begin{eqnarray}\label{general}
&& \bra{\psi}(A_{1}B_{1}- A_{1}B_{-1}+ A_{-1}B_{-1}-
A_{-1}B_{1})\ket\psi  \nonumber \\ && = \bra{\psi}(A_{1}-
A_{-1})(B_{1} - B_{-1})\ket\psi  =  \inpr{A}{B},
\end{eqnarray}
where $\ket{A}\equiv (A_{1}- A_{-1})\ket{\psi}$ and $\ket{B}\equiv
(B_{1}- B_{-1})\ket{\psi}$. So for the purpose of calculating
expectation values the generalized observables ${\cal A}, {\cal
B}, \tilde{a}, \tilde{b}$ can each be represented by a single
hermitian operator, namely $(A_{1}- A_{-1})$, $(B_{1}- B_{-1})$,
$(a_{1}- a_{-1})$ and $(b_{1}- b_{-1})$, respectively; the joint
measurements are represented by the corresponding products.
Compared to the case discussed in sect.\ \ref{cirelson}, the
differences are that the operators $A_{i}, B_{j}, \ldots$ need not
be projection operators, and the squares of $(A_{1}- A_{-1}),
(B_{1}- B_{-1}), \ldots$ need not be $\openone$. The second proof
of the Cirel'son inequality given in section \ref{cirelson} goes
nevertheless through, because the operators $(A_{1}- A_{-1}),
(B_{1}- B_{-1}), \dots$ all have norms $\leq 1$.

Indeed,
\begin{eqnarray}\label{norm1}
&&||(A_{1}- A_{-1})\ket{\psi}||^{2} \nonumber \\
&& = ||A_{1}\ket{\psi}||^{2}+ ||A_{-1}\ket{\psi}||^{2}
-2\inpr{A_{1}\psi}{A_{-1}\psi},
\end{eqnarray}
whereas \begin{eqnarray}\label{norm2}  &&||(A_{1}+
A_{-1})\ket{\psi}||^{2} \nonumber \\ && = ||A_{1}\ket{\psi}||^{2}+
||A_{-1}\ket{\psi}||^{2} +2\inpr{A_{1}\psi}{A_{-1}\psi}=1,
\end{eqnarray}
so that
\begin{equation}\label{norm}
||(A_{1}- A_{-1})\ket{\psi}||^{2}= 1 -
4\inpr{A_{1}\psi}{A_{-1}\psi}.
\end{equation}
Because $A_{-1}=\openone - A_{1}$, $[A_{1}, A_{-1}]=0$ and the
inner products in (\ref{norm1}), (\ref{norm2}) and (\ref{norm})
are real. This inner product is also $\geq 0$:
\begin{equation}\label{inpr}
\inpr{A_{1}\psi}{A_{-1}\psi}= \bra{\psi}A_{1}\ket{\psi} -
\bra{\psi}A_{1}^{2}\ket{\psi},
\end{equation}
which is $\geq 0$ because $A_{1}$ has norm $\leq 1$ and only has
eigenvalues $\lambda_{i}$ with $ 0 \leq \lambda_{i} \leq 1$.

Now introduce vectors $\ket{a}, \ket{b}$ in the obvious way:
$\ket{a}\equiv(a_{1}-
a_{-1})\ket{\psi},\ket{b}=(b_{1}-b_{-1})\ket{\psi}$. It follows
from the above that the vectors $\ket{A}, \ket{B}, \ket{a},
\ket{b}$ all have norms $\leq 1$, just as the vectors denoted by
the same symbols in sect.\ \ref{cirelson}. Repeating the proof
(\ref{cirel}), we therefore find:
\begin{eqnarray}\label{cirgen}
  &&|\langle {\cal A}{\cal B} + {\cal A} \tilde{b} + \tilde{a} {\cal B} -
  \tilde{a}\tilde{b}\rangle|  \nonumber \\ && = |\inpr{A}{B+b} + \inpr{a}{B-b}| \leq
  2\sqrt{2}.
\end{eqnarray}
This inequality holds in every pure state $\ket{\psi}$. Its
validity in any mixed state $\rho$ follows immediately.

\section{The strongest inequality}\label{stronger}

Cirel'son's inequality is not the only nor the strongest one that
can be derived from the locality (no-signaling) and therefore
commutativity requirement. Put $X \equiv \langle {\cal A}\tilde{b}
+ \tilde{a} {\cal B}\rangle$ and $Y \equiv \langle {\cal A} {\cal
B} - \tilde{a}\tilde{b}\rangle$. Cirel'son's inequality can now be
written as
\begin{equation}\label{general1}
  |X + Y| \leq 2\sqrt{2}.
\end{equation}
In the $X, Y$ `correlation plane' this inequality restricts the
points $(X,Y)$ to the strip between the two lines
\begin{equation}\label{lines1}
X + Y = \pm 2\sqrt{2}.
\end{equation}
But by a minimal change in the proof of sect.\ \ref{cirelson} it
immediately follows that also the following inequality holds:
\begin{equation}\label{general2}
  |X - Y| \leq 2\sqrt{2},
\end{equation}
so that the points must also lie in the strip bounded by the lines
\begin{equation}\label{lines2}
X - Y = \pm 2\sqrt{2}.
\end{equation}
We also have the obvious inequalities $|X| \leq 2, \;\; |Y| \leq 2
$, so that the allowed points $(X,Y)$ must be in the intersection
of the interiors of the two squares indicated in Fig.\ 1.
\begin{figure}
\begin{picture}(150,150)
\put(75,75){\circle{40}} \put(55,95){\line(1,0){40}}
\put(55,55){\line(1,0){40}} \put(55,55){\line(0,1){40}}
\put(95,95){\line(0,-1){40}} \put(103.3,75){\line(-1,1){28.3}}
\put(103.3,75){\line(-1,-1){28.3}} \put(46.7,75){\line(1,1){28.3}}
\put(46.7,75){\line(1,-1){28.3}} \put(0,75){\line(1,0){150}}
\put(75,0){\line(0,1){150}} \put(145,69){\makebox(0,0){$X$}}
\put(69,145){\makebox(0,0){$Y$}}
\end{picture}
\caption{The $X, Y$ plane. The slanted square represents
inequalities (\ref{general1}) plus (\ref{general2}), the circle
inequality (\ref{strongest}).}
\end{figure}
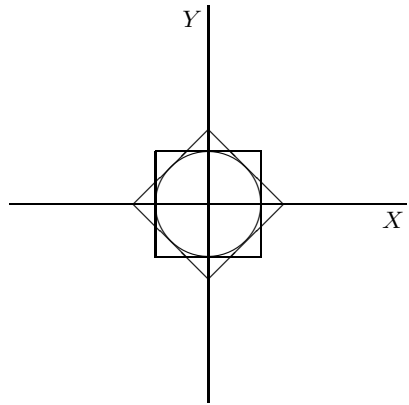
It further turns out that they must be inside (or on the sides of)
all squares that result from these just-mentioned squares by
applying an arbitrary rotation around an axis through the origin
of the $X, Y$ plane and normal to this plane. To prove this,
consider the expression $| X\sin\varphi + Y\cos\varphi|$. We have:
\begin{eqnarray}\label{general3}
  &&|X\sin\varphi + Y\cos\varphi| \nonumber \\
  &=& |\inpr{A}{B\cos\varphi + b\sin\varphi} + \inpr{a}{B\sin\varphi -
  b\cos\varphi}| \nonumber \\
  & \leq & ||\, \ket{B\cos\varphi} + \ket{b\sin\varphi} || + ||\, \ket{B\sin\varphi} - \ket{
  b\cos\varphi}|| \nonumber \\
  & \leq &  \sqrt{\sin^{2}\varphi + \cos^{2}\varphi +
  2\mbox{\bf Re}\inpr{B}{b}\sin\varphi\cos\varphi} \nonumber\\ && + \: \sqrt{\sin^{2}\varphi + \cos^{2}\varphi -
  2\mbox{\bf Re}\inpr{B}{b}\sin\varphi\cos\varphi} \nonumber \\
  & \leq &  \sqrt{\sin^{2}\varphi + \cos^{2}\varphi} + \sqrt{\sin^{2}\varphi +
  \cos^{2}\varphi}= 2.
\end{eqnarray}
Cirel'son's inequality and the other inequalities mentioned
earlier in this section are special cases of this general set of
inequalities (in which $\varphi$ can take arbitrary values). It
should be noted that these proofs apply both to the case of
ordinary observables and to the case of generalized observables.

It is clear from the geometry of Fig.\ 1 that the requirement that
all the inequalities (\ref{general3}) be satisfied leads to the
inequality
\begin{equation}\label{strongest}
  X^{2} + Y^{2}= \langle {\cal A}\tilde{b} + \tilde{a} {\cal
B}\rangle^{2} + \langle {\cal A} {\cal B} -
\tilde{a}\tilde{b}\rangle^{2} \leq 4.
\end{equation}
All points $X, Y$ are inside or on the circumference of a circle
with radius $2$.

Inequality (\ref{strongest}) (which was recently proved directly,
by a variational argument, for the case of ordinary spin
observables by Uffink \cite{uffink}) summarizes all generalized
Cirel'son inequalities (\ref{general3}). All values $X, Y$ that
satisfy (\ref{strongest}) also satisfy all Cirel'son inequalities
(\ref{general3}); but satisfaction of a finite number of
inequalities of (\ref{general3}) is not sufficient to guarantee
satisfaction of (\ref{strongest}). Moreover, each point on the
circumference of the circle can actually be attained, because the
bound of the corresponding generalized Cirel'son inequality can be
attained (the one resulting in a line tangent to the circle in the
point in question). Inequality (\ref{strongest}) is therefore the
strongest inequality in terms of $X, Y$ that follows from the
requirement of commutativity.

\section{Locality}\label{locality}

Bell's inequality (\ref{bell}) is valid for an arbitrary quadruple
of stochastic functions on one probability space, and as such is
not immediately connected with locality issues. The link with
locality comes in via the application of (\ref{bell}) to
situations of the Einstein-Podolsky-Rosen type in which A and a,
and B and b, stand for measurements on the space-like separated
wings 1 and 2 of the experiment, respectively. An experimenter at
1 can choose between measuring A and a; her or his colleague at 2
has the choice between B and b. The combined measurement on parts
1 and 2 is represented by the product of the individual single
system result functions. This is justified by a locality
assumption: a measurement of a physical quantity on one wing of
the experiment has no influence at the other wing. On the basis of
this assumption one-wing quantities are represented by one and the
same function, regardless of whether, and if so which, measurement
is performed at the other side. Both the possible measurement
outcomes and their probabilities are insensitive to choices made
at the other side.

An obvious silent assumption in this is that $A, a, B, b$
correspond to characteristic measurement devices and interactions.
The device corresponding to $A$, e.g., should remain the same in
different instances---that the possible outcomes and corresponding
probabilities remain the same is by itself not enough. Consider to
make this clear a Stern-Gerlach device at wing 1 that undergoes a
rotation depending on the choice between $B$ and $b$: although the
possible outcomes would still be $+1$ and $-1$ and the
probabilities would remain equal to $1/2$ (if the device measures
the spin of a spin-1/2 particle), this would not constitute one
specific measured quantity. Spin along different axes would be
measured. In spite of the fact that all measurement results could
be represented by the same function $A$, the rotation of the
corresponding device would signal non-locality. Within the quantum
formalism such a non-invariance of the measuring procedure could
easily lead to a violation of (\ref{strongest}). An explicit
example can be constructed by stipulating that the concrete
physical implementation of measuring ${\cal A} {\cal B}$ and the
other joint quantities in (\ref{strongest}), in the two-particle
singlet state, be: ``Measure $\sigma_{x}^{I}$ and
$-\sigma_{x}^{II}$, call the results `spin of particle $I$ along
axis $\alpha$ and spin of particle $II$ along axis $\beta$,
respectively'; perform the same measurement of $\sigma_{x}^{I}$
and $-\sigma_{x}^{II}$ to obtain the spin values for the pairs of
axes $\alpha^{\prime},\beta$ and $\alpha,\beta^{\prime}$, and
measure $\sigma_{x}^{I}$ and $+\sigma_{x}^{II}$ in the case of
$\alpha^{\prime},\beta^{\prime}$''. Obviously, the correlation
functions obtained in this way violate Cirel'son's inequality
maximally (even though outcomes and probabilities at each wing are
insensitive to what choice is made at the other wing). This is
because the quantities defined in this way are not bona-fide local
physical quantities in the sense we have discussed. Indeed,
according to this measurement protocol the measuring procedure for
the spin of particle $II$ along $\beta^{\prime}$ depends on
whether spin along $\alpha$ or along $\alpha^{\prime}$ is measured
on particle $I$.

So we have identified a first locality requirement: the operators
that are used to represent physical quantities on the individual
wings of the experiment should refer to the same physical devices
and interactions, regardless of what goes on at the other side. It
should be possible to measure these quantities on wing 1 and wing
2, respectively, together; therefore the operators representing
them should be compatible (i.e., commuting or jointly measurable
in the sense of the POVM scheme).

The second locality assumption to be considered is that no signals
are transmitted: the possible outcomes and their probabilities on
the two sides of the experiment are insensitive to what happens at
the other wing. Usually, this is the only assumption that is
explicitly discussed.

Within the orthodox treatment of measurements in quantum theory
the compatibility assumption and the no-signaling assumption are
equivalent: both lead to the requirement that the operators at one
side commute with those at the other side. This commutativity is
in turn sufficient to derive Cirel'son's inequality and its
generalizations, including inequality (\ref{strongest}). Within
the framework of generalized measurements it is the no-signaling
requirement that leads to commutativity; so here we have to invoke
locality (in the sense of no signaling, or relativistic causality
in the relativistic context) explicitly in order to derive
(\ref{strongest}); the requirement of joint measurability is not
enough.

The two just-mentioned locality requirements together lead to a
theoretical description with one-wing operators (referring to
invariant measuring procedures) that commute with those at the
other wing. This is sufficient for deriving the inequalities of
section \ref{stronger}. These inequalities can therefore be used
in experimental tests of locality. If an inequality is violated in
experiments, this indicates either the propagation of influences
that change the outcomes and/or their probabilities (an
application of this idea can be found in \cite{beckman}), or it
indicates non-invariance of the measured quantity (see above for
an example of the latter possibility).\footnote{This conclusion
assumes the exclusive use of the ordinary (predictive) quantum
rule for computing expectation values that says that the state of
the system is represented by a density operator $\rho$ and that
$\langle A\rangle=\tr\{\rho A\}$. If one uses ensembles that
cannot be represented by a quantum state, for example if one
calculates averages in post-selected ensembles, the situation
becomes very different. As Cabello has shown recently
\cite{cabello}, Cirel'son's inequality can no longer be derived if
combinations of different post-selected ensembles are used for the
calculation of average values.}

\section{An argument by Popescu and Rohrlich}\label{popescu}
The conclusion of the previous Section is that quantum mechanical
locality (no-signaling) is responsible for the existence of the
upper bound of Cirel'son's inequality. This conclusion might seem
in conflict with an argument by Popescu and Rohrlich \cite{popescu
and rohrlich}. These authors argue that the impossibility of
signaling does \emph{not} limit the sum of the correlations
occurring in Cirel'son's inequality to $2\sqrt{2}$. Their
counterexample is an EPR situation in which spin measurements are
performed on the two wings. For the outcomes of these measurements
a particular probability distribution is postulated, as follows.
The two possible outcomes are taken to be $+1$ and $-1$ along any
axis, and both of these possibilities are postulated to have a
probability of $1/2$, independently of the measurement performed
at the other wing. So the measured outcomes and their
probabilities do not give any information about choices made at
the other side; it is impossible to signal, or, as Popescu and
Rohrlich put it, relativistic causality is satisfied.

Further, for any pair of axes, the combinations of outcomes $+1,
+1$ and $-1, -1$ are assumed to be equally probable, and the same
applies to the combinations $+1, -1$ and $-1, +1$. Finally, the
correlation function (a `superquantum' correlation function) is
stipulated to have a form like the following:
  \begin{equation}
  E(\theta) = \left\{ \begin{array}{ll}
   +1 & \mbox{for $0 \leq \theta \leq \pi/4$}    \\
   2 - 4\theta/\pi & \mbox{for $\pi /4 \leq \theta \leq 3\pi/4$} \\
   -1  & \mbox{for $3\pi/4 \leq \theta \leq \pi$}
  \end{array} \right.\label{supercorr}
  \end{equation}
This is equivalent to assuming that the probability
$p_{++}(\theta)$ of the pair of outcomes $+1,-1$ is given by
\[p_{++}(\theta) = \frac{E(\theta) + 1}{4}.\] In these formulas
$\theta$ is the angle between the axes on the left and right,
respectively, along which the spin measurements are made.

Now consider four axes $\alpha^{\prime}, \beta, \alpha,
\beta^{\prime}$ separated by successive angles of $\pi/4$ and
lying in one plane. We find that
\begin{equation}
 E(\alpha,\beta)+ E(\alpha^{\prime},\beta) +
 E(\alpha,\beta^{\prime})- E(\alpha^{\prime},\beta^{\prime}) =
 4. \label{violation}
 \end{equation}
Cirel'son's inequality can therefore be violated, even to the
maximum extent logically possible, by a correlation function that
respects invariance of one-wing outcomes and probabilities, and
thus the impossibility of signaling.

Clearly, therefore, the requirement that outcomes and
probabilities are invariant is by itself insufficient to derive
Cirel'son's inequality and its generalizations. Nevertheless, we
have demonstrated above that Cirel'son's inequality \emph{can} be
derived from that locality assumption within the theoretical
framework of quantum mechanics. It is only when the no-signaling
requirement is implemented within a well-defined theoretical
framework, equipped with prescriptions for how to represent
observables (corresponding to characteristic measuring
procedures), that it gets enough bite to make the derivation of
Bell-type inequalities possible. Indeed, we have seen that within
the framework of theories that operate with a phase space on which
physical quantities are represented by (stochastic) functions, the
original Bell inequalities can be obtained, whereas within the
Hilbert space formalism of quantum theory inequality
(\ref{strongest}) results. In both cases the correlation function
postulated by Popescu and Rohrlich cannot arise in a local way (it
\emph{can} be produced by non-local means as was illustrated
earlier in this section).

The fact that within the framework of quantum theory the
correlation function (\ref{supercorr}) cannot be produced in a
local way (because it violates (\ref{strongest})) does not mean,
of course, that there cannot exist other theoretical frameworks in
which `superquantum' correlation functions \emph{could} arise in a
local way; frameworks that use neither functions on a state space
nor the Hilbert space operator formalism. It is difficult to say
anything definite about such  hypothetical theoretical frameworks.
Popescu's and Rohrlich's argument does not use any assumption
about how the observables are represented mathematically, and
therefore does not provide a sufficient basis for a discussion of
locality and causality (in fact, we already observed that their
probability distribution could be produced by non-local
mechanisms, which implies that invariance of outcomes and
probabilities is a necessary but not a sufficient condition for
causality).

Summing up, Popescu and Rohrlich are right in their claim that
Bell-type inequalities do not follow from the no-signaling
requirement alone. But this does not answer the question of
whether locality, in the sense of the no-signaling requirement,
may lead to such inequalities \emph{in the context of specific
theories}. It turns out that if locality is fleshed out within
classical theories, this leads to Bell's inequality; if it is
fleshed out within the framework of quantum theory this leads to
inequality (\ref{strongest}).

\begin{acknowledgments}
I thank Jos Uffink for helpful remarks.
\end{acknowledgments}


\begin{thebibliography}{99}
\bibitem{cirelson}
B.S. Cirel'son, Lett.\ Math.\ Phys.\ {\bf 4}, 93 (1980).
\bibitem{landau}
L.J. Landau, Phys.\ Lett.\ {\bf A 120}, 52 (1987).
\bibitem{popescu and rohrlich}
S. Popescu and D. Rohrlich, Found.\ Phys.\ {\bf24}, 379
(1994); \\
quant-ph/9508009; quant-ph/9605004; quant-ph/9709026.
\bibitem{ludwig}
G. Ludwig, \emph{Einf\"{u}hrung in die Grundlagen der
Theoretischen Physik}, Vol.\ 3 (Vieweg, Braunschweig, 1976).
\bibitem{davies}
E.B. Davies, \emph{Quantum Theory of Open Systems} (Academic
Press, New York, 1976).
\bibitem{uffink0}
J. Uffink, Int.\ Jnl.\ Theor.\ Phys.\ {\bf33}, 199 (1994).
\bibitem{busch}
P. Busch and J. Singh, Phys.\ Lett.\ A {\bf249}, 10 (1998).
\bibitem{uffink}
J. Uffink, Phys.\ Rev.\ Lett.\ {\bf88}, 230406 (2002).
\bibitem{beckman}
D. Beckman, D. Gottesman, M.A. Nielsen, and J. Preskill, Phys.\
Rev.\ A {\bf64}, 052309 (2001).
\bibitem{cabello}
A. Cabello, Phys.\ Rev.\ Lett.\ {\bf88}, 060403 (2002).
\end{thebibliography}
\end{document}